\documentclass[aps,prl,twocolumn,showpacs,superscriptaddress,groupedaddress]{revtex4}  
\usepackage{graphicx}  
\usepackage{dcolumn}   
\usepackage{bm}        
\usepackage{amssymb}   
\usepackage{color}
\usepackage{gensymb}

\usepackage[normalem]{ulem}
\def\={\,=\,}

\newcommand{\Lagr}{\mathcal{L}}

\begin{document}

\title{Sub-Kelvin Lateral Thermal Transport in Diffusive Graphene}

\affiliation{Department of Physics, Duke University, Durham, NC 27708, USA}
\affiliation{Oak Ridge National Laboratory, Oak Ridge, TN 37830, USA}
\affiliation{Department of Physics, City University of Hong Kong, Kowloon, Hong Kong SAR}
\affiliation{Department of Physics and Astronomy, Appalachian State University, Boone, NC 28607, USA}
\author{A. W.~Draelos} \affiliation{Department of Physics, Duke University, Durham, NC 27708, USA}
\author{A.~Silverman} \affiliation{Department of Physics, Duke University, Durham, NC 27708, USA}
\author{B.~Eniwaye} \affiliation{Department of Physics, Duke University, Durham, NC 27708, USA}
\author{E. G.~Arnault} \affiliation{Department of Physics, Duke University, Durham, NC 27708, USA}
\author{C. T.~Ke} \affiliation{Department of Physics, Duke University, Durham, NC 27708, USA}
\author{M. T.~Wei} \affiliation{Department of Physics, Duke University, Durham, NC 27708, USA}
\author{I.~Vlassiouk} \affiliation{Oak Ridge National Laboratory, Oak Ridge, TN 37830, USA}
\author{I. V. Borzenets} \affiliation{Department of Physics, City University of Hong Kong, Kowloon, Hong Kong SAR}
\author{F.~Amet} \affiliation{Department of Physics and Astronomy, Appalachian State University, Boone, NC 28607, USA}
\author{G.~Finkelstein} \affiliation{Department of Physics, Duke University, Durham, NC 27708, USA}

\date{\today}

\begin{abstract}
In this work, we report on hot carrier diffusion in graphene across large enough length scales that the carriers are not thermalized across the crystal. The carriers are injected into graphene at one site and their thermal transport is studied as a function of applied power and distance from the heating source, up to tens of micrometers away. Superconducting contacts prevent out-diffusion of hot carriers to isolate the electron-phonon coupling as the sole channel for thermal relaxation. As local thermometers, we use the amplitude of the Universal Conductance Fluctuations, which varies monotonically as a function of temperature. By measuring the electron temperature simultaneously along the length we observe a thermal gradient which results from the competition between electron-phonon cooling and lateral heat flow. 
\end{abstract}

\pacs{}
\maketitle

Graphene exhibits a unique combination of a small electron heat capacitance, weak electron-phonon coupling strength, and high electrical and thermal conductivity \cite{castro,nika,nika2}. These features make graphene an ideal candidate for bolometric applications, such as the detection of microwave and terahertz radiation \cite{mckitterick2,santavicca, Yan2012, Machida2014}. The thermal properties of graphene have been extensively studied in the low-temperature regime, where the weak coupling between electrons and acoustic phonons becomes the dominant cooling pathway for hot electrons \cite{Hwang, Kubakaddi, Bistritzer, Viljas, Voutilainen2011}.

At low temperatures, the electron-phonon cooling power density depends on the electron and phonon temperatures, $T_e$ and $T_{ph}$, as: $\dot{q}_{ep}  = \Sigma$ $(T_e^{\delta} - T_{ph}^{\delta})$, where $\Sigma$ is the electron-phonon coupling strength per unit area. Earlier experimental work showed that $\delta$ was typically 4 in clean graphene \cite{Efetov2010, Voutilainen2011, Price2012, betz, mckitterick} as predicted by theory \cite{Kubakaddi, Bistritzer, Viljas}.  However, $\delta$ can be reduced to 3 due to supercollisions and disorder-assisted scattering \cite{chen, song, McEuen2012, betz2, Borzenets, somphonsane, Eless2013, Han2013, Laitinen2014, Vora, Multiterminal}. 

Previous work showed this power law relationship throughout the sample, indicating the electrons were well-thermalized spatially across distances of several micrometers \cite{fong2,mckitterick}. However, lateral temperature gradients can develop in larger devices due to an interplay between hot electron diffusion, phonon thermal conductivity, and local electron-phonon cooling \cite{betz,fong1, fong2,mckitterick,zuev,wei,Sierra2015,checkelsky}. Here we use a large enough graphene crystal to explore the interplay between the local cooling and the lateral thermal conductance within the graphene crystal. We use superconducting metal contacts that prevent the outflow of hot electrons to the leads, which allows us to study the thermal pathways within the graphene itself.

\begin{figure}
\includegraphics[width=3.1in]{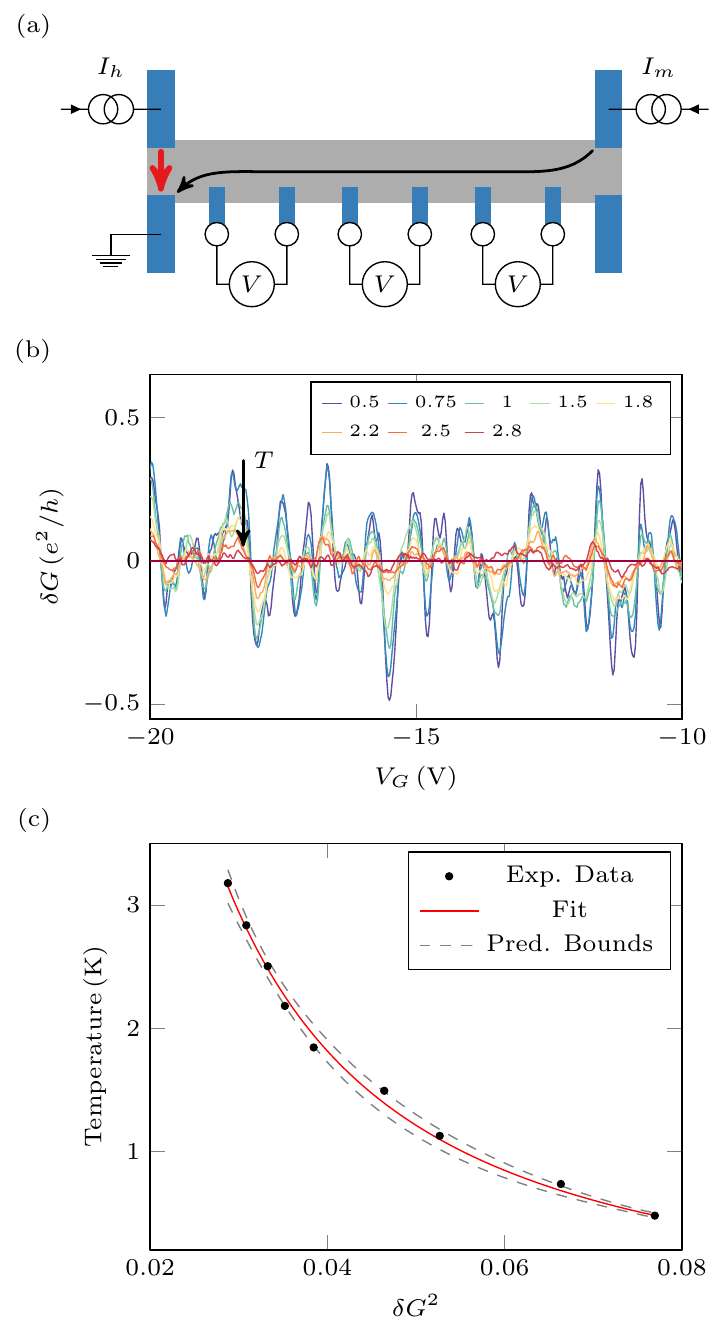}%
\caption{\label{fig:epsart} (a) Diagram of the sample showing the heating current $I_h$, measurement current $I_m$, and voltage probes $V$ placed along one side of the graphene device. Heating on the left edge of the sample allows us to probe how the heat is transferred over to the right edge. (b) A calibration curve of the Universal Conductance Fluctuations ($\delta $G) as a function of back gate voltage ($V_{G}$) at different sample temperatures. As the temperature (in K) increases the repeatable conductance fluctuations decay and the overall variance decreases. (c) Data (black dots) and polynomial fitted curve (red line) to establish the one-to-one correspondence between temperature and the variance of the conductance for a particular gate voltage to calibrate a single junction. The grey dashed lines are the 95\% prediction bounds associated with the calibration fit.}
\end{figure}

Studying temperature gradients in a long strip of graphene at low temperatures provides insights into the competition between local heat flow from the hot electrons to the phonon baths, $\dot{q}_{ep}$, and lateral Wiedemann-Franz heat diffusion, $\dot{q}_{WF}$. In the stationary regime, for a given supplied power density $p(x)$:
\begin{eqnarray}
p(x)=\dot{q}_{WF} + \dot{q}_{ep}  =  \Sigma (T^{\delta} - T_{ph}^{\delta} ) - \Lagr \sigma \nabla \cdot (T \cdot \nabla T) 
\label{eq:one}
\end{eqnarray}
where $\Lagr$ is the Lorenz number ($\pi^2 k_B^2 / 3e^2$) and $\sigma$ is the electrical conductivity in the graphene. 
The typical scale over which a temperature gradient could develop can be estimated by comparing the two cooling power terms: $a \propto \sqrt{\Lagr \sigma/2\Sigma T_{ph}}$. At 1 K, we can expect to see a temperature gradient begin to develop over about 5 $\mu$m. At lower temperatures, this gradient is even larger and thus not noticeable in typically-sized samples ($< 10$ $\mu$m).

We therefore fabricated a sample with a long strip (5 x 50 $\mu$m) of large-domain ($\approx 100 ~ \mu$m) CVD monolayer graphene \cite{vlassiouk}. The graphene layers were detached from the growth substrate using ammonium persulfate copper etchant and transferred onto a p-doped Si substrate with a 300 nm thermal oxide that served as a gate \cite{petrone}. Lead-indium superconducting contacts with a critical temperature $T_c \sim 7 $ K and a superconducting gap $\Delta \sim 1.1 $ meV \cite{jeong} were placed along the length of the graphene (see Fig. 1). The contacts only overlap with the edges of the graphene, in order not to interfere with the diffusion of hot electrons down the length of the strip. Each pair of contacts are spaced by a large enough distance ( $> 5$ $\mu$m) that the graphene is not proximitized and no supercurrent is observed. (See supplementary for fabrication details). 

To measure the local electron temperature, we used the variance of the Universal Conductance Fluctuations (UCF) measured in junctions formed by pairs of contacts along the length of the sample \cite{kechedzhi,bora,somphonsane}. To calibrate these thermometers, we first uniformly heated the entire sample with an external source (on the sample holder, well thermalized to the sample substrate). For a given substrate temperature, the conductance profile for each junction vs. gate voltage was measured in the four-probe setup while sending a small AC current of $\sim$ 10 nA across the entire length of graphene (Figure 1a). 

Figure 1b shows the evolution of the resulting fluctuations $\delta$G with temperature for one of the junctions; they decay controllably with increasing T.  The rich physics of these fluctuations does not concern us here; instead, the UCF average amplitude was analyzed to produce a calibration curve characteristic for each junction and gate voltage range. The UCF variance is thus correlated to the known substrate temperature (Fig 1b,c). In the lowest temperature range (T $<$ 0.5 K) this curve saturates as the conductance fluctuations are no longer sensitive to the temperature, and their variance reaches an upper bound. We consequently operate our dilution refrigerator at reduced capacity with the base temperature of the sample holder of 0.5 K. We limit the upper bound of temperature to 3 K, which is below the $T_c$ of the superconducting contacts, in order to avoid out-diffusion of hot electrons to the leads.

The graphene was then locally heated at the base temperature of 0.5 K by passing a relatively large current $I_h$ = 0.1-10 $\mu$A between a pair of heater contacts on one end of the strip. The voltage drop across this heater was simultaneously measured to determine the power applied in the form of Joule heating, $P=VI$. This method directly creates hot electrons in the graphene, instead of using a closely-spaced separate heater that would require heat to be transported through the substrate before reaching the graphene. A smaller measurement current of 10 nA was passed along the length of the sample in the same manner as it was done during the UCF calibration. It produces negligible heat compared to the heating current. The calibrated variance of the UCF for each junction was then used to extract the local electron temperature for a given applied current. 

Figure 2 shows the evolution of the local electron temperature as a function of the heating power and position for the carrier density of approximately 2-3 $\times$10$^{12}$ cm$^{-2}$. Initially well-thermalized at the sample holder temperature of 0.5 K, the thermal distribution gradually becomes non-uniform at high power, as junctions closer to the heater become significantly hotter. 

The problem is effectively one-dimensional since the width of the sample (5 $\mu$m) is small enough for electrons to be efficiently thermalized across the sample. Note that $T_{ph}$ here is not negligible compared to the electron temperature $T_e$. Equation (1) can be simplified if we define $y$ as the square of the dimensionless temperature $(T_e/T_{ph})^2$, and the characteristic length scale $a= \sqrt{\Lagr \sigma/2\Sigma T_{ph}^{\delta-2}}$:

\begin{equation}
\frac{d^{2}y}{dx^{2}} = \frac{1}{a^{2}}(y^{\delta/2} - 1) - \frac{2 p(x)}{\Lagr \sigma}
\label{eq:three},
\end{equation}

\begin{figure}
\includegraphics[width=3.25in]{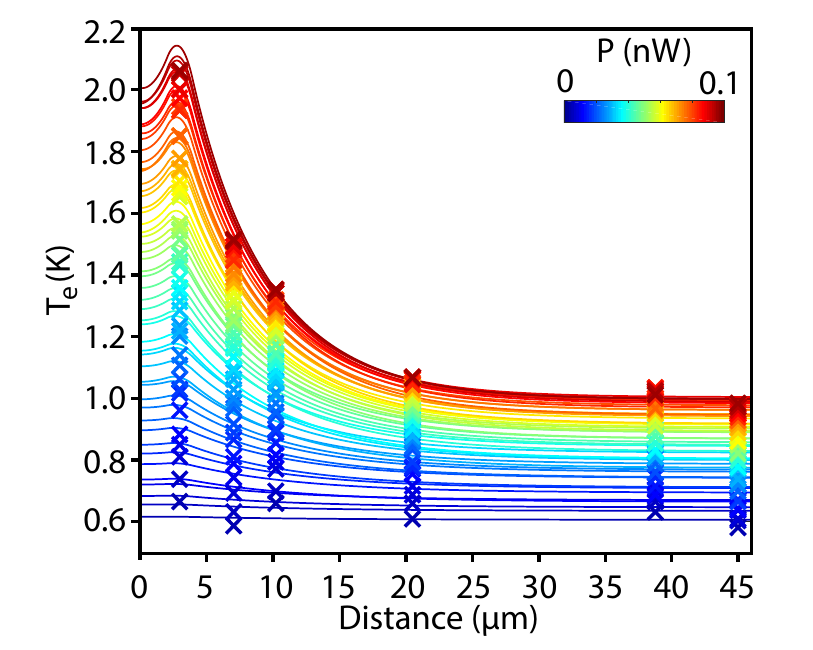}%
\caption{\label{fig:wide} Electron temperature versus distance from the heating source at different applied powers (colorbar). A non-uniform gradient is seen clearly developing at increasing heating powers. Continuous curves correspond to solutions of equation (2) for powers ranging from 0 to 0.1nW, with $\delta\,=\,4$ and $\Sigma\,=\,$0.32 W K$^{-4}$ m$^{-2}$. The phonon temperature $T_{ph}$ is a fitting parameter and is taken to be equal to the average temperature at the end of the strip. The heater is located between x$\,=\,$2 $\mu$m and x$\,=\,$4 $\mu$m, which explains the position of the temperature maximum.}
\end{figure}

The power density $p(x)$ is assumed to be constant around the heater and zero elsewhere. Additionally, the heat flow $\Lagr \sigma T\nabla T$ must vanish at both ends of the strip, which yields the two boundary conditions $y'(0)\,=\,y'(L)\,=\,0$. This differential equation is solvable numerically, using the electron phonon coupling $\Sigma$ and the phonon temperature $T_{ph}$ as fitting parameters. However, in practice $a$ is much smaller than the total length of the strip so the electron temperature is expected to reach $T_{ph}$ at the far end of the strip. We therefore take $T_{ph}$$\,=\,$$T_e$(L) and fit each curve with the single parameter $\Sigma$. Our results are best fit by $\delta=4$ and $\Sigma \approx 0.32 \pm .09$ W K$^{-4}$ m$^{-2}$ (See supplementary). This exponent corresponds to the electron cooling rate in the clean case; indeed the electron temperature range in this experiment is higher than or comparable to the characteristic temperature scale introduced by the disorder \cite{chen, mckitterick}. This temperature was estimated to be $T_{dis}=\frac{30\hbar s\zeta(3)}{\pi^{4}k_{B}l_{mfp}}$, where $l_{mfp}$ and $s$ are the mean free path and phonon velocity respectively. Our estimated mean free path of 50 nm yields a crossover temperature of about 1 K.

The temperature profiles solutions of (2) for each total power are plotted as continuous curves on Figure 2 and are in good agreement with our data. Note that this value of the electron-phonon coupling corresponds to a length scale $a\,\approx\, $ 8 $\mu$m for the temperature range $T_{ph} = 0.5 - 1 $ K: this is in good agreement with our observations and justifies the approximation $T_e(L) \approx T_{ph}$ for $L\gg a$. Using the theoretical expression for $\Sigma$, we find that a deformation potential D $\approx$ 36 eV, in agreement with prior experimental works \cite{fong2}.

\begin{figure}
\includegraphics[width=3.5in]{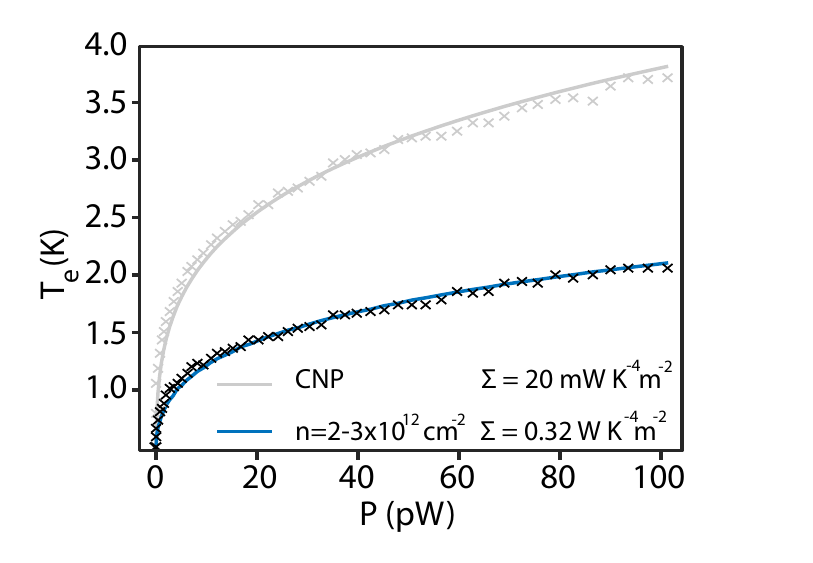}%
\caption{\label{fig:std} Electron temperature versus applied heating power at x=0$\mu$m around charge neutrality (gray) and for an electron density of 2-3$\times$ 10$^{-12}$cm$^{-2}$. The electron phonon coupling constant $\Sigma$ is heavily suppressed close to charge neutrality. 
}
\end{figure}

Interestingly, the measured electron temperature rises significantly higher than the base temperature, even tens of microns away from the heater (Figure 2). The hot electron diffusion decays on the scale $a \ll L$, so $T(L)$ is therefore expected to asymptotically approach the phonon temperature. Therefore, it appears that both electrons and phonons at the far end of graphene reach temperature higher than that of the substrate. (By measuring a test sample which had both the heater and the thermometer on the same chip, we have checked that the applied power on the scale of $P \lesssim 100$ pW causes negligible rise of the substrate temperature; see supplementary.) Overheating of phonons in graphene is possible if they do not efficiently couple to the substrate. Little is known about phonon coupling between graphene and the substrate at $T\sim1$ K, and further studies of the observed behavior are clearly needed. 

Finally, we return to the electrons and measure their temperature close to the Dirac point. In this regime, the electron-phonon coupling is expected to drop due to the reduced electron phase space. The lateral heat conductivity should also be reduced due to lower electronic conductivity. Both factors should contribute to the increase in the equilibrium electron temperature close to the heater, as in this case the applied heat dissipates less efficiently. Figure 3 shows $T_e$ vs. $p$ in the junction closest to the heater in two cases: around charge neutrality (top curve) and at 2-3 $\times 10^{-12}$ cm$^{-2}$ (bottom curve, same density as in Fig. 2). Both fits are obtained assuming $\delta\,=\,4$. At charge neutrality, $\Sigma$ is found to be $\sim 20 \,$mW K$^{-4}$ m$^{-2}$. Unfortunately, our method of extracting the local temperature from averaging the UCF over 10 V of gate voltage prevents us from meaningfully extracting the density dependence beyond affirming the expected strong suppression of $\Sigma$ around charge neutrality.


	In summary, we have studied the spatially non-uniform distribution of electron temperatures in a large graphene sample, at temperatures of the order of 1 K. We measured the local temperature as a function of distance from the source of heating, which allowed us to explore the interplay between the lateral electron diffusion and the local cooling electron by phonons. A simple modeling allows us to fit the experimental results with a realistic electron-phonon coupling constant. Finally, we observe an intriguing rise of electron temperature far beyond the region directly affected by heating. 


Transport measurements conducted by A.W.D., E.G.A., and G.F. were supported by Division of Materials Sciences and Engineering, Office of Basic Energy Sciences, U.S. Department of Energy, under Award No. DE- SC0002765. Lithographic fabrication and characterization of the samples performed by A.W.D. and A.S. and B.E. was supported by ARO Award W911NF16-1-0122 and NSF awards ECCS-1610213 and DMR-1743907. F.A. was supported by Army Research Office (award No W911NF-16-1-0132) and the North Carolina Space grant (award No 2015-1942-AP-05). I.V.B. acknowledges CityU New Research Initiatives/Infrastructure Support from Central (APRC): 9610395.

\pagebreak
\begin{center}
\textbf{\large Supplementary to: Sub-Kelvin Lateral Thermal Transport in Diffusive Graphene}
\end{center}

\setcounter{figure}{0}
\renewcommand{\thefigure}{S\arabic{figure}}

\section{Device Fabrication}

CVD (chemical vapor deposition)-grown graphene is used for producing large scale films required for this study. The large-domain monolayer graphene is grown on a copper substrate, and then transferred to the Si/SiO$_2$ silicon substrate which is used as a back gate. To transfer, first a polymer (PMMA, poly(methyl methacrylate)) layer is spin-coated onto one surface of the copper to protect the graphene. The other side of the copper undergoes an O$_2$ plasma ashing for 30 s at 90 W RF power to fully expose the copper underneath. The copper is then placed exposed-side down into an ammonium persulfate (APS) chemical bath to be etched away\cite{Suk}, followed by a rinse in de-ionized (DI) water. The polymer layer floats and supports the graphene as the copper is etched, leaving a PMMA/graphene film on the surface of the last water bath. The silicon substrate is then placed in the water and used to pick up the polymer/graphene film at a $\sim 45\degree$ angle to minimize the residual water. The transferred film is baked on a hot plate at $150\degree$C to remove wrinkles and water residue. Once the film has fully dried, the polymer supporting layer was then dissolved in DCM (dichloromethane) to reveal the complete transferred graphene layer.

The sample was patterned into the rectangular shape for thermal measurements (5 x 50 $\mu m$) with standard electron beam lithography (EBL) techniques, using PMMA as the electron resist. For our device, we defined a pattern using an NPGS system \cite{Nabity} equipped FEI XL 30 SEM (scanning electron microscope). Multiple stages of EBL were used to make a single device. The first stage is to deposit a design of gold bonding pads and leads that will interface with the graphene region. During this stage a grid of small gold markers are also written. These serve as alignment markers for subsequent EBL steps that require finer resolution. Second, an etch write is done to define the shape of the sample. Oxygen plasma (30 s) is sufficient to etch away any unwanted graphene not covered by PMMA. Finally, a third stage of EBL is used to pattern the superconducting leads directly connecting to the sample. This stage is left until the end, as the low melting point ($\sim150\degree$C) of the lead-indium (PbIn) prohibits additional lithography steps due to the high temperature needed to cure the PMMA. 

\begin{figure}
\includegraphics[width=3.15in]{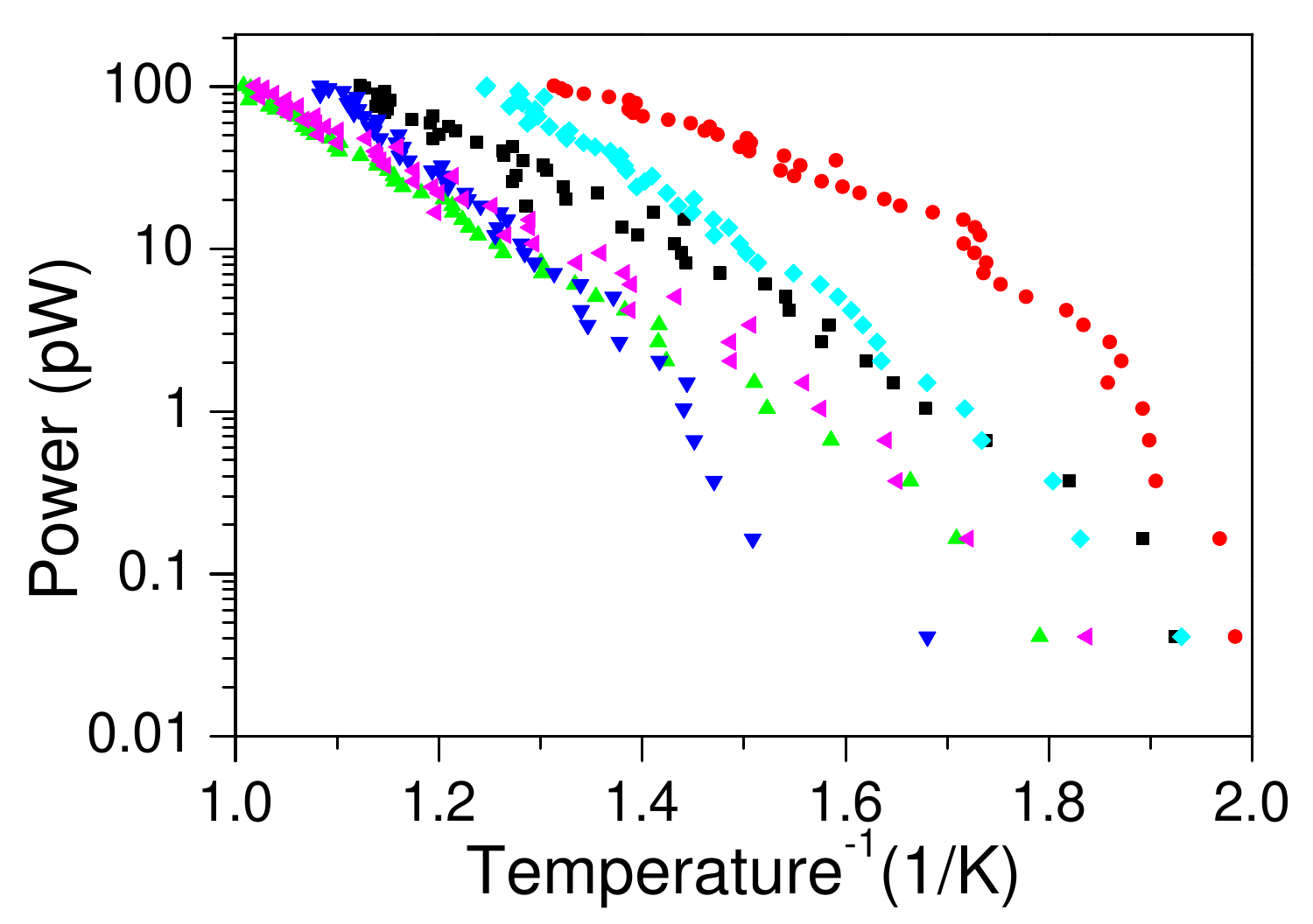}%
\caption{\label{fig:std} (a) Applied heating power vs inverse temperature at the far end of the strip $T^{-1}(L)$, assumed to be equal to the phonon temperature $T_{ph}$. Data is shown for different values of gate voltage $V_G$: $-20V$ (black), $-10V$ (red), $0V$ (green), $10V$ (blue), $20V$ (light blue), $30V$ (magenta).  Power is plotted on a logarithmic scale. }
\end{figure}
\begin{figure}
\includegraphics[width=3.15in]{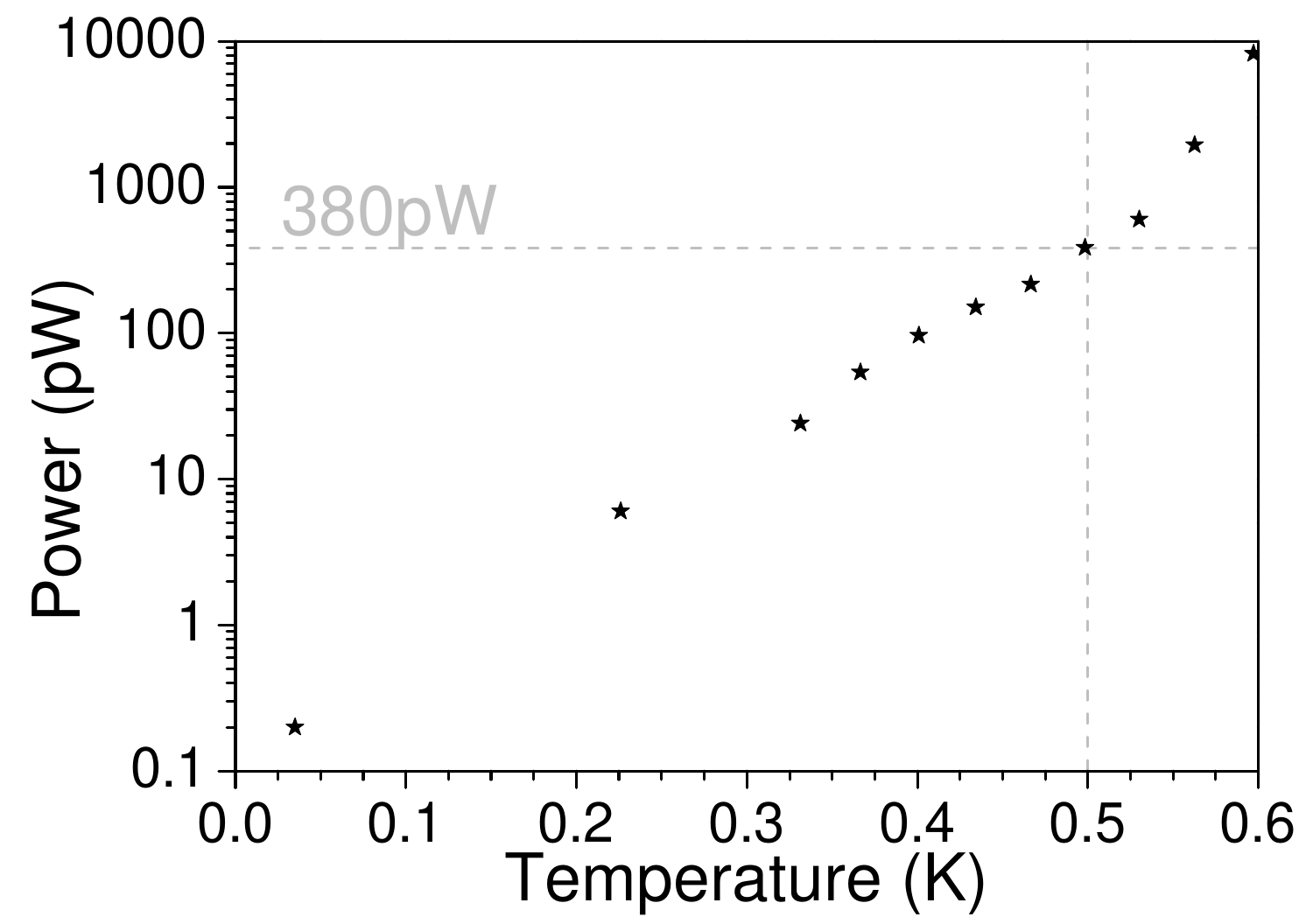}%
\caption{\label{fig:std} Heating power versus temperature for the case where the graphene strip acting as a heater, and the strip acting as a thermometer are separated by $3\mu m$. In this case, heat exchange can only be mediated via the Si/SiO$_2$ substrate.  }
\end{figure}

Finally, lead-indium (PbIn) superconducting contacts were evaporated onto the device as described in the main text. Building upon our previous work with lead contacts, here we used instead an alloy of lead and indium \cite{Jeong}. Pb oxidizes rapidly upon exposure to air, which severely degrades the contacts. Combining the Pb with In reduces the metal's oxidation, without any significant reduction in the critical temperature. We first create the alloy by melting both materials together in a vacuum deposition chamber. Holding the temperature of the crucible above their melting points and below the evaporation point for 10 min allows the metals to intermix before being co-evaporated. Roughly 100 nm of this PbIn alloy is deposited at a high rate of 2 nm/s and moderate vacuum (x$10^{-5}$ mbar) to ensure small metal grain sizes. LN$_2$ is used to keep the substrate cooled. 

\begin{figure*}
\includegraphics[width=6.0in]{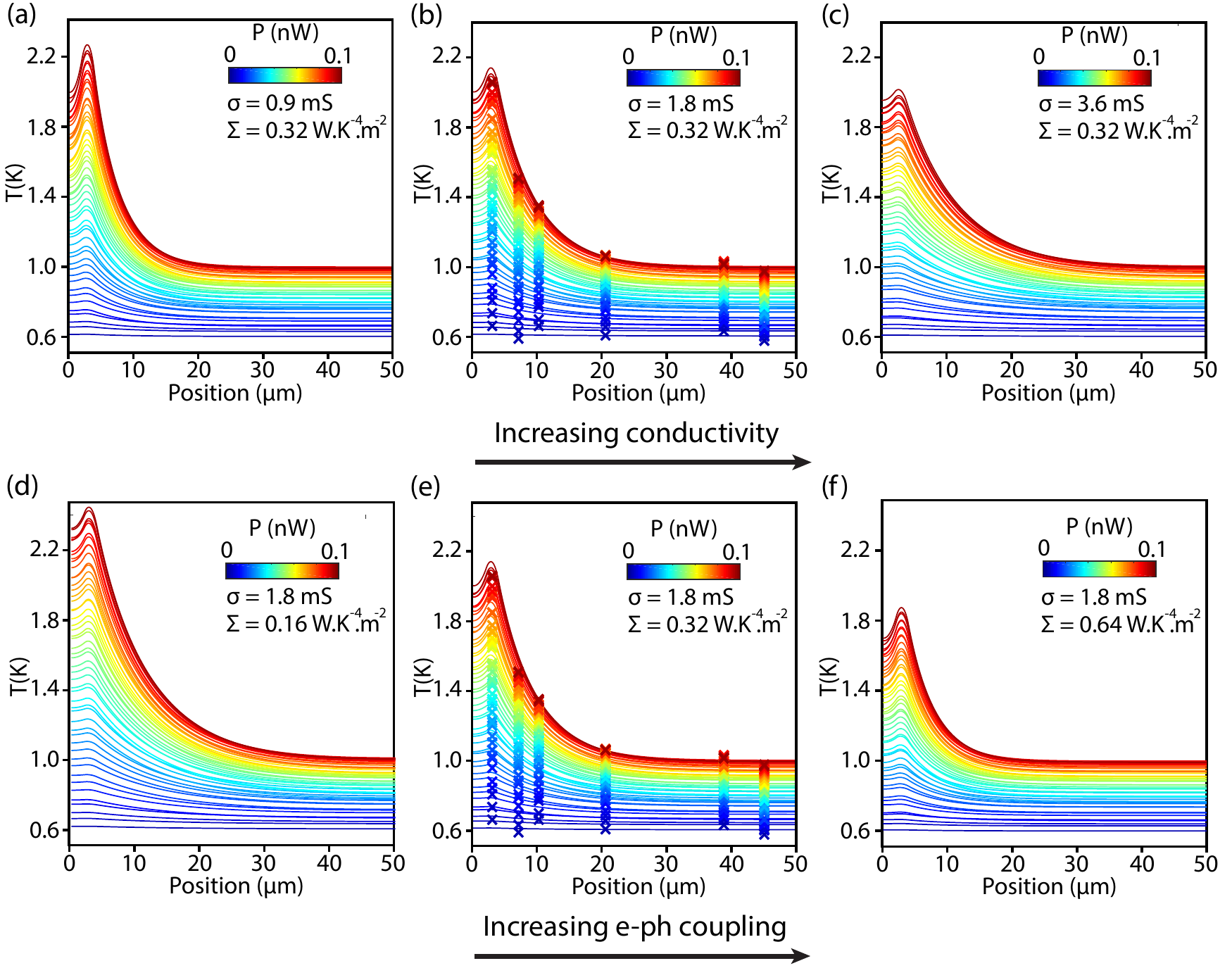}%
\caption{\label{fig:std} Evolution of the temperature profile as a function of electrical conductivity and electron phonon coupling. (a-c) Temperature profiles for the electrical conductivities of $\sigma\,=\,0.9$ mS, 1.8 mS, and 3.6 mS. The electron phonon coupling is held constant at 0.32 W.m$^{-2}$.K$^{-4}$. Panel (b) is identical to figure 2 of the main paper and shows actual data points for convenience. (d-f) Temperature profiles for an electron-phonon coupling strength of $\Sigma$ = 0.16, $\Sigma$ = 0.32 and $\Sigma$ = 0.64  W.m$^{-2}$.K$^{-4}$. The electrical conductivity is held constant at 1.8mS. Panel (e) is identical to panel (b) duplicated for convenience.}
\end{figure*}

\begin{figure}
\includegraphics[width=3.0in]{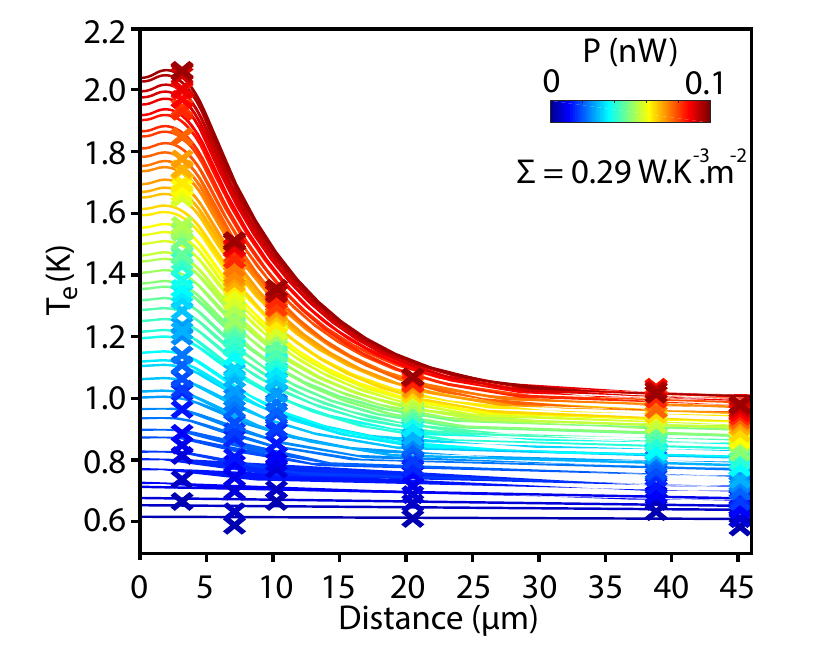}%
\caption{\label{fig:std} Electron temperature versus distance from the heating source at different applied powers. Continuous curves correspond to solution of the heat equation (2) for powers ranging from 0 to 0.1 nW. Here, we assume the disordered case with $\delta\,=\,3$, and find that the fit quality is worse than with $\delta\,=\,4$.
}
\end{figure}

\section{Characterization of Substrate Heating}

As explained  in the main paper, the temperature increases significantly at the far end of the graphene strip, $T(L)$, which we attribute to an increase in the phonon temperature $T_{ph}$ in graphene. Figure S1a shows $T(L)$ as a function of the applied power. The cooling power at the far end of the strip appears to follow a faster than power-law dependence (Fig. S1). This dependence, as well as the lack of trend in the gate voltage, are presently not understood.

To quantify the effects of heating on the Si/SiO$_2$ substrate, we used a different sample that also combined the heater and thermometer. In brief, we find the local heating of the substrate (and the sample holder) is negligible at the heating power applied in this paper. 



Two graphene Josephson junctions separated by  about 3 $\mu$m were fabricated on top of the same type of Si/SiO$_2$ substrate as studied in the main text. One of the junctions served as a thermometer, and another as a heater. First, we measured the critical current $I_C$ of the thermometer junction as a function of the overall sample holder temperature, as controlled by a global heater and a resistance thermometer mounted on the sample holder. Next, with the global heater off, heating current $I_H$ was applied to the heater junction. Critical current of the thermometer junction $I_C$ was measured as a function of the Joule heating power $P=I_H^2 R_H$. (Here, $R_H$ is the resistance of the heater.) Thus, a curve of temperature vs heating power T(P) could be calculated (Figure S2). (For detailed methodology see Refs. \onlinecite{Multiterminal, Phonon}). 

We see that when the two devices are only connected via the substrate we require a heating power of $P\sim400$ pW to reach a temperature of $500$mK (Figure S2). Returning to the sample studied in the main text, we conclude that the rise of the substrate temperature should be negligible, and it does not explain the increased electron temperature at distances of a few tens of microns from the heater.


\section{Modeling}

The local electron temperature $T(x)$ along the graphene strip is expected to solve the stationary non-linear heat equation with local heating $P$ and electron-phonon coupling $\Sigma$, all expressed per unit of area:

\begin{equation}
0=P+\Sigma (T^{\delta} - T_{ph}^{\delta} ) - \Lagr \sigma \nabla \cdot (T \cdot \nabla T) 
\end{equation}

Here $T_{ph}$ represents the phonon temperature, $\sigma$ the electrical conductivity, $\Lagr$ the Lorenz number and $\delta$ an exponent. 

A change of variable $y=(T/T_{ph})^{2}$ and the definition of the length scale $a= \sqrt{\Lagr \sigma/2\Sigma T_{ph}^{\delta-2}}$ allows us to rewrite the differential equation as:

\begin{equation}
\frac{d^{2}y}{dx^{2}} = \frac{1}{a^{2}}(y^{\delta/2} - 1) - \frac{2 p(x)}{\Lagr \sigma}
\end{equation}

As stated previously, the vanishing heat flow at $x=0$ and $x=L$ yields the boundary conditions $y'(0)=y'(L)=0$, with L=50 $\mu$m. We approximate the Joule heating power density $p(x)$ as constant and only finite between x=2 $\mu$m and x=4 $\mu$m which corresponds to the extent of the heating contacts.

We then solve the differential equation iteratively for y'(L)=0 and a dense array of trial values for y(L); the final solution y(x) is then the one that verifies $y'(0)=0$ as well. 

In order to illustrate trends in the temperature profile, in Figure S3 we present solutions of the heat equation for different values of the electrical conductivity and the electron-phonon coupling $\Sigma$. As the electrical conductivity increases, heat diffusion through the electron bath is facilitated, which results in a temperature increase farther from the source, and a shallower temperature gradient close to it. When the electron-phonon coupling $\Sigma$ increases, the local electron temperature is of course lower and decays to $T_{ph}$ much faster.

In order to get an estimate of the electron phonon coupling $\Sigma$, we generate the temperature profile $T(x)$ for an array of values of $\Sigma$ and use a least square fitting procedure to extract $\Sigma\approx 0.32\pm.09 $ W.m$^{-2}$.K$^{-4}$

Finally, for disordered graphene, the cooling power of phonons is enhanced and its scaling with temperature has an exponent $\delta\,=\,3$. As explained in the main text, the crossover temperature to that regime is expected to be on the order of 1 K. Figure S4 shows a fit to our data using solutions of Equation (2) with $\delta\,=\,3$, with the optimal fit found for $\Sigma\,=\,0.29$ W.K$^{-3}$.m$^{-2}$. As expected the temperature gradient for a given $T$ is a little less steep than for $\delta\,=\,4$. Overall, we conclude that $\delta\,=\,4$ describes our data better than $\delta\,=\,3$.

\end{document}